\documentclass{article}

\usepackage[english]{babel}
\pdfoutput=1
\usepackage[a4paper,top=2cm,bottom=2cm,left=3cm,right=3cm,marginparwidth=1.75cm]{geometry}
\usepackage{amsmath,amssymb,amsfonts}
\usepackage{algorithmic}
\usepackage{graphicx}
\usepackage{float}
\usepackage{caption}
\usepackage{subcaption}
\usepackage{amsmath}
\usepackage{tabularx}
\usepackage{adjustbox}
\usepackage{tcolorbox}
\usepackage{textcomp}
\usepackage{xcolor}
\usepackage{color, colortbl}
\usepackage{colortbl}
\usepackage{siunitx}
\PassOptionsToPackage{hyphens}{url}\usepackage{hyperref}
\usepackage{cleveref}
\usepackage[utf8]{inputenc}
\usepackage{csquotes}
\usepackage{booktabs}
\usepackage[numbers]{natbib}
\usepackage{longtable}
\usepackage{array}
\usepackage{url}
\usepackage{titlesec}
\usepackage{authblk}
\renewcommand{\thetable}{\Roman{table}}

\titleformat{\subsection}
  {\mdseries\itshape\large} 
  {\thesubsection}{1em}{} 

\usepackage[english]{babel}

\crefformat{figure}{#2Figure~#1#3}
\Crefformat{figure}{#2Figure~#1#3}
\crefformat{table}{#2Table~#1#3}
\Crefformat{table}{#2Table~#1#3}
\crefformat{section}{#2Section~#1#3}
\Crefformat{section}{#2Section~#1#3}

\author[1]{Faheem Ullah}
\author[2]{Xiaohan Ye}
\author[3]{Uswa Fatima}
\author[4]{Zahid Akhtar}
\author[2]{Yuxi Wu}
\author[2]{Hussain Ahmad}

\affil[1]{Zayed University, United Arab Emirates}
\affil[2]{The University of Adelaide, Australia}
\affil[3]{National University of Science and Technology, Pakistan}
\affil[4]{State University of New York Polytechnic Institute, USA}

\title{What Skills Do Cyber Security Professionals Need?}
\date{}

\begin{document}
\maketitle

\begin{abstract}
\textbf{Purpose} - The increasing number of cyber-attacks has elevated the importance of cybersecurity for organizations. This has also increased the demand for professionals with the necessary skills to protect these organizations. As a result, many individuals are looking to enter the field of cybersecurity. However, there is a lack of clear understanding of the skills required for a successful career in this field. In this paper, we identify the skills required for cybersecurity professionals. We also determine how the demand for cyber skills relates to various cyber roles such as security analyst and security architect. Furthermore, we identify the programming languages that are important for cybersecurity professionals. 

\textbf{Design/Methodology} - For this study, we have collected and analyzed data from 12,161 job ads and 49,002 Stack Overflow posts. By examining this, we identified patterns and trends related to skill requirements, role-specific demands, and programming languages in cybersecurity.

\textbf{Findings} -  Our results reveal that (i) communication skills and project management skills are the most important soft skills, (ii) as compared to soft skills, the demand for technical skills varies more across various cyber roles, and (iii) Java is the most commonly used programming language.

\textbf{Originality} - Our findings serve as a guideline for individuals aiming to get into the field of cybersecurity. Moreover, our findings are useful in terms of informing educational institutes to teach the correct set of skills to students doing degrees in cybersecurity. 
\\
\\
\textbf{Keywords:} Cybersecurity; Security Workforce; Data Analysis; Stack Overflow; SoC; Incident Response; Threat Detection; Risk Management
\end{abstract}

\section{Introduction}
\label{sec:Intro}
\subsection{Background and Motivation}
The cybersecurity landscape is evolving rapidly, driven by the growing sophistication of cyber threats and the continuous advancement of technology. This dynamic environment demands that cybersecurity professionals adapt and expand their skills to address increasingly complex challenges. Central to this evolution are modern software paradigms such as microservice architectures \cite{jayalath2024microservice, abdulsatar2024towards, ahmad2024smart, ahmad2024towards}, Large Language Models (LLMs) \cite{chopra2024chatnvd, haque2022think}, and Large Concept Models (LCMs) \cite{ahmad2025future}, the cybersecurity landscape has entered a new era of both opportunity and challenge. As technology continues to advance and the threat of cybercrime grows more sophisticated \cite{goel2024optimizing, goel2024machine, goel2023enhancing}, organizations are in greater need of a highly skilled workforce to help defend against these threats \cite{37}. Cybercrime is estimated to cost around \$9.22 trillion in 2024 and is expected to escalate to \$17.9 trillion by 2030, nearly doubling during this period \cite{ahmad2024survey}. Despite this, there remains a significant shortage of skilled cybersecurity professionals, a trend that is expected to persist in the coming years. Samantha cites the current cybersecurity talent crunch as a lack of adequate cyber skills understanding \cite{36,35}. In its 2022 Cybersecurity Workforce Study, ((ISC)²) proposes that the global cybersecurity workforce will grow by 11.1\% in 2022 compared to 2019 \cite{31}. The survey revealed that nearly 70\% of security professionals in multiple areas do not believe their organization has enough staff to be effective \cite{31}. 
\smallbreak
While the global shortage of cybersecurity professionals is increasing, so is the threat and sophistication of cybercrime \cite{33}. From 2020 to 2025, Cybersecurity Ventures predicts that global losses caused by cybercrime will grow at an annual rate of 15\% resulting in \$10.5 trillion losses by 2025 \cite{25}. With cybercrime on the rise, businesses, and governments need a skilled cybersecurity workforce to build cyberspace protection \cite{34}. With this in mind, it is essential that individuals looking to enter the field of cybersecurity understand the skills and competencies that are in demand.
\smallbreak
Knowing which cybersecurity skills are in high demand is important for a number of reasons. Firstly, it helps individuals who are looking to enter the field of cybersecurity to focus their efforts and resources on the most in-demand skills \cite{27}. Secondly, it helps educational institutes to better align their curriculum with the needs of the industry. By offering courses that focus on the most in-demand skills, they can help their students to better prepare for the job market \cite{conklin2014re}. Thirdly, for businesses and organizations, knowing which skills are in high demand can help them to find the right talent for their cybersecurity needs \cite{30}. Finally, knowing which skills are in high demand can also help governments to develop policies and programs that better support the growth and development of the cybersecurity workforce \cite{cobb2016mind}.
\smallbreak
While it is important to identify the skills required for cybersecurity professionals, developing a comprehensive list of the skills required for cybersecurity professionals is difficult for several reasons. First, the field of cybersecurity is constantly evolving \cite{furstenau202020}. With the advancement of technology and the emergence of new threats \cite{ahmad2023review}, the skills required for cybersecurity professionals are continually changing. Second, different organizations have different requirements for cybersecurity professionals \cite{10,12}. For example, a large financial institution may require more expertise in network security, while a small startup may focus more on data privacy and protection. Thirdly, the cybersecurity field is not well regulated, which means there are no standardized training programs or certifications that guarantee a certain level of proficiency \cite{srinivas2019government}. This further complicates the task of developing a comprehensive list of skills required for cybersecurity professionals.
\smallbreak
Given the need to understand key cyber skills, we answer multiple Research Questions (RQ) in this paper to identify the key cyber skills, the relation of those skills with various cyber roles, and the associated programming languages. Our study is based on the collection and analysis of 12,161 job ads and 49,002 Stack Overflow posts. These job ads were collected from various job portals that cover the global cybersecurity market.  

\subsection{Research Questions}
In this study, we conducted a large-scale study of Stack Overflow and job advertisements to understand skills that security professionals need to learn and practice by answering the following research questions.
\smallbreak
\textbf{RQ1: What skills do security professionals need?} Answering this question will help individuals and organizations to build a strong and effective cybersecurity workforce to protect against the growing threat of cybercrime. Understanding the expertise needed also guides educational institutes in teaching the right skills to future cybersecurity professionals.
\smallbreak
\textbf{RQ2: Whether the demand for skills vary across different cybersecurity roles?} The cybersecurity industry has different roles such as security analyst, security architect, and consultant, etc. Some roles are more technical compared to others. We explore the relationship between the required skills and cyber roles. 
\smallbreak
\textbf{RQ3: What programming languages do security professionals need to use?} Several cyber roles require a thorough understanding and practice of programming languages in their daily jobs. Understanding the programming languages in demand can help security professionals in skill development and career advancement within the cybersecurity field.

\subsection{Contributions}
To summarize, we make the following contributions in this paper:

\begin{itemize}

    \item We have performed a thorough analysis of 12,161 cybersecurity job ads and 49,002 security-related posts on Stack Overflow. 
    \item We have identified 6 soft skills, 20 hard skills, and 13 professional certifications that are important for cybersecurity professionals.
    \item We have identified 7 programming languages, along with their respective importance, for cybersecurity professionals.
    \item Finally, we have studied the relationship between various skills and cybersecurity roles to determine whether or not various cyber roles require specific skills. 
\end{itemize}

The structure of this paper is as follows: Section \ref{RW} positions the novelty of our work with respect to the related works. Section \ref{RM} describes the research methodology followed to collect and analyze data from job ads and Stack Overflow to answer the RQs. Section \ref{R} presents our findings with respect to the three RQs. Section \ref{D} reports the implications and threats to the validity of our findings. Finally, Section \ref{C} concludes the paper.

\section{Related Work}\label{RW}

Several studies have been conducted on the identification of skills that are critical for various domains such as cybersecurity (e.g., \cite{2,3,5,10,12,15,29,caulkins2019cybersecurity,jerman2022cybersecurity,peslak2019cybersecurity, jain2024comprehensive}), software engineering (e.g., \cite{khaouja2021survey,daneva2019understanding,matturro2019systematic, goel2018overview, goel2021maintenance}), and signal processing \cite{abbas2024robust}. In the context of cybersecurity, Ben-Asher and Gonzalez \cite{2} analyzed data collected from a case study with 55 participants in the US to examine how individuals with and without security knowledge detect malicious events. Unlike this study, our study is based on global data collected from job ads and Stack Overflow. Also, our research questions are different. In \cite{3}, the authors conducted a review of the literature to identify the gaps in cybersecurity expertise. Then, this paper emphasizes the contribution of social fit in a highly complex and heterogeneous cyber workforce. Although this paper's aim is similar to our first RQ, our work differs from it in terms of the data source, two RQs, and the findings. Parker and Brown et al., \cite{10} analyzed data from 196 job ads in South Africa to determine the skills required for cybersecurity professionals. In \cite{peslak2019cybersecurity}, the authors collected data from 500 job ads to identify critical cyber skills. Similar to our study,  \cite{10} and \cite{peslak2019cybersecurity} collected data from job ads. However, these papers collected data only from 150 - 500 job ads in one country. In comparison, we have collected data from 12, 161 jobs from around the world and 49, 002 Stack Overflow posts. 
\smallbreak
Potter and Vickers et al. \cite{12} collected data from interviews with security professionals to identify the skill critical for security professionals in the Australian market. This study differs from ours in terms of the RQs and data analyzed. In \cite{15}, the authors focussed on skills important for one particular cyber role - information security analyst. This study was conducted based on a literature review in the Malaysian context. Our study is not limited to one particular cyber role. Chowdhury and Gkioulos \cite{29} carried out a literature review of existing studies to identify the skills that security professionals need to secure critical infrastructures. Unlike our study which identifies cyber skills for general cybersecurity, Chowdhury and Gkioulos \cite{29} only focused on the skills required for one domain i.e., critical infrastructures. Caulkins et al. \cite{caulkins2019cybersecurity} conducted a survey with government cybersecurity professionals to identify the soft skills required for cybersecurity professionals. Unlike \cite{caulkins2019cybersecurity}, our study focuses not only on soft skills but also hard skills and certifications. Jerman et al, \cite{jerman2022cybersecurity} conducted interviews with students, teachers, and parents to understand what cybersecurity topics should be taught to the students and how best they can be taught. Jones et al., \cite{5} conducted 44 interviews with cybersecurity professionals to identify cyber topics that students should learn in school. Unlike \cite{5} and \cite{jerman2022cybersecurity} which focus on cyber topics, our study focuses on cyber skills. In summary, our work differs from the existing works in terms of RQs, data source, and findings.

\section{Methodology}\label{RM}
\begin{figure}[ht]
 \centering
 \makebox[\textwidth][c]{\includegraphics[width=1\textwidth]{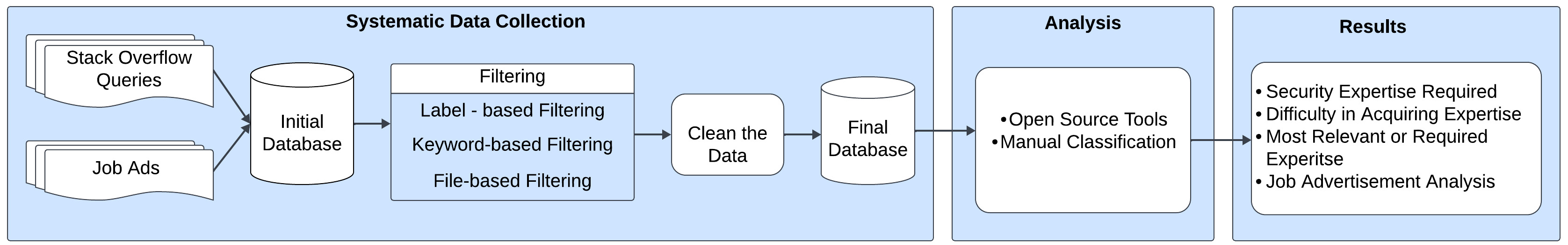}}%
 \caption{Research methodology used for conducting this study}
 \label{method}
\end{figure}

The methodology used for conducting this study is illustrated in Figure \ref{method}. As shown, data was collected from Stack Overflow and Job Ads. 

\subsection{Stack Overflow}
\begin{figure}[ht]
 \centering
 \makebox[\textwidth][c]{\includegraphics[width=11cm, height=2.5cm]{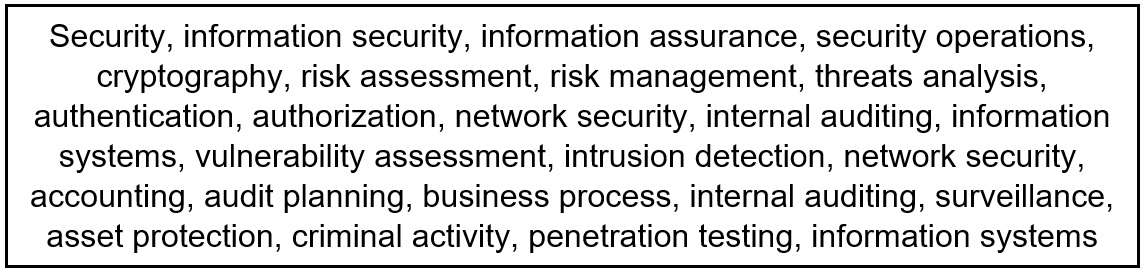}}%
 \caption{Terms used for searching Stack Overflow posts.}
 \label{search_terms}
\end{figure}

Stack Overflow\footnote{https://stackoverflow.com/} is a highly regarded platform for those in the computing industry, providing a space for academic exchange and solutions to daily problems faced by programmers. The platform is widely used by developers and security professionals as a means of learning and understanding complex expertise. To understand recent developments and trends in cybersecurity skills, the query function in Stack Overflow's Stack Exchange Data Explorer  was (SEDE)\footnote{https://data.stackexchange.com/stackoverflow/query/new} utilized by writing SQL code. Our data collection period for security skills spanned the past five years, starting on June 22 and ending on July 27, 2022.
\smallbreak
\textbf{\textit{Data Collection and Pre-processing:}} The metadata for this study was collected from various sources, including web searches, CyberSeek\footnote{https://www.cyberseek.org/index.html}, and relevant literature \cite{2,3,5,10,12,15,29,caulkins2019cybersecurity,jerman2022cybersecurity,peslak2019cybersecurity}. A comprehensive set of search terms was compiled, which are listed in Figure \ref{search_terms}. We used these search terms to collect data from Stack Overflow. \textit{Data Pre-processing:} After the data was collected, data pre-processing was performed by retaining only the ‘search terms’ and ‘score’ portions of the data to ensure the accuracy and efficiency of data analysis. This was done using a Python script in Jupyter Notebook, with the help of the NLTK \footnote{https://www.nltk.org/} library, which was used to remove stop words and distinguish different tags. \textit{Data Analysis and Visualization:} We utilized open-source exploratory data analysis tools from GitHub to analyze the data obtained from Stack Overflow repositories. To make the results more understandable, the data was manually integrated and visualized using Excel.

\begin{figure}[ht]
 \centering
 \makebox[\textwidth][c]{\includegraphics[width=11cm, height=3cm]{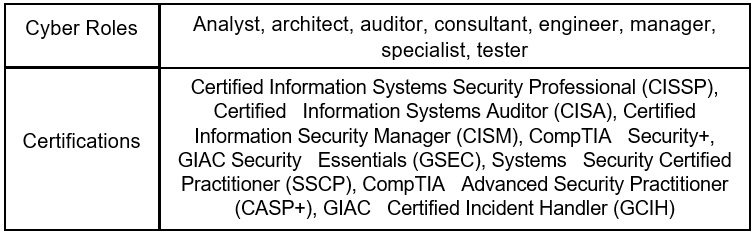}}%
 \caption{Search terms for cyber roles and certifications}
 \label{search_terms_roles}
 \vspace{-0.5 cm}
\end{figure}

\subsection{Job Ads}
To determine how cybersecurity-related skills and tools are used in practice, we collected and analyzed data based on job advertisements from job portals, including Dice\footnote{https://www.dice.com/}, 
SA Job Search\footnote{https://jobsearch.gov.au/}, and
Seek\footnote{https://www.seek.com.au/}.
\smallbreak
\textbf{\textit{Keyword-based data acquisition.}} Jupyter Notebook and a Python script were utilized to perform the initial filtering of job advertisements based on relevant keywords and positions. The keywords and positions used for the filtering process were sourced from CyberSeek. These keywords are presented in Figure \ref{search_terms_roles}. The script then extracted the job title, recruitment link, recruiter, and job description of each filtered advertisement. The critical information was then organized into a CSV file. This data was collected from June 29 to July 10, 2022, resulting in a total of 12,161 job advertisement data points. The results are presented in Table \ref{data}.

\begin{table}[h]
	\caption{The number of job ads collected from different sources and the number of posts collected from Stack Overflow.}\label{data}
	\centering
	\scriptsize
	\begin{tabular}{|p{4cm} |p{4cm}|}
		\hline
		\textbf{Data Source}     &\textbf{Number of Job Ads/posts}         \\\hline\hline
		Seek                    &4643 \\
		DICE                    &3700 \\ 
		Jora                    &1552 \\
		SA Job Search           &2266 \\ 
		Total Job Ads           &12161 \\\hline
		Total Stack Overflow Posts                   &49002 \\ \hline
	\end{tabular}
	\vspace{-0.5 em}
\end{table}
\smallbreak
\textbf{\textit{Data pre-processing}} To streamline the data analysis process and improve its efficiency, it was necessary to clean up the data and eliminate any redundant information. The data pre-processing process involved using a Python script that utilizes the NLTK library to retain the job descriptions, eliminate stop words and remove any punctuation. The cleaned data was then analyzed using the advertools \footnote{https://github.com/eliasdabbas/advertools} library to calculate keyword frequency and produce a new, organized CSV file. To further simplify the data, we utilized the advertools library to rank the keywords by frequency, with a minimum phrase length of 2. Some non-technical terms were still present in the analyzed keywords, so we organized the technical terms into Excel to merge and sort the data from the three recruitment websites by position, and visualized the results. The formula for calculating the keyword frequency is shown in Equation \ref{equation_1}, where k$_{i}$ (i = 1, 2, 3, 4) represents the corresponding number of keywords for different recruitment websites.

\begin{equation}
    KeywordFrequency = \sum_{i=1}^{n}k_{i} \qquad
    \label{equation_1}
\end{equation}

\textbf{\textit{Manual classification}} To ensure a thorough analysis of the technical terms, the relevant keywords were organized into an Excel spreadsheet. This allowed us to integrate and sort the data from the three recruitment websites by position and visually represent the findings. As the study utilized two data sources, only the data from Stack Overflow allowed to freely choose the time frame. On the other hand, job advertisement data was updated on an annual basis and any job postings that successfully recruited the required personnel would be deleted. The choice to collect data from Stack Overflow over a five-year period was made to strike a balance between the relevance and persuasiveness of the data. If the time frame was too long, the risk of including irrelevant or outdated information increased. Conversely, if the time frame was too short, the data would only reflect the current skill needs of the cybersecurity industry and lack long-term perspectives. The process is illustrated in Figure \ref{method}.

\section{Results}\label{R}
In this section, we report our findings with respect to the three research questions presented in Section \ref{sec:Intro}. 

\subsection{RQ1: Security skills required}

We analyzed the identified skills from three different perspectives - soft skills, hard skills, and professional certifications. skill is considered as a soft skill if it is hard to quantify and is important for working effectively with others. On the other hand, skill is considered as a hard skill if it is easy to quantify and is crucial for executing technical tasks. Professional certifications are often a requirement for a job as is mentioned in the job ad. These certifications are issued by well-known entities such as CISCO.  
\smallbreak
Table \ref{soft_skills} presents the most crucial soft skills for cybersecurity professionals. According to our results, communication and project management are the two most important soft skills for security professionals. Strong communication skills are critical for multiple reasons. First, security professionals often work in teams to tackle complex security threats. Second, clear and concise communication skills help security professionals to coordinate an effective response and minimize damage in case of an attack. Third, security professionals need to communicate with a wide range of stakeholders including executives, IT teams, and end-users. Similarly, project management skills are crucial for security professionals because they help to effectively plan, manage, and execute complex security projects. Project management skill also helps to effectively adapt to the changing cyberspace and adjust project plans as needed.
\smallbreak
\begin{table}[t]
\caption{Identified soft skills for cybersecurity professionals. The numbers in the table denote the number of job ads/posts mentioning the particular skill.}\label{soft_skills}
\scriptsize
\resizebox{\textwidth}{!}{%
\begin{tabular}{|l|l|l|l|l|l|l|}
\hline
\textbf{Rank} & \textbf{Skill} & \textbf{Dice} & \textbf{SA Job Search} & \textbf{Seek} & \textbf{Stack Overflow} & \textbf{Total} \\ \hline \hline
1 & Communication & 862 & 20 & 157 & 39 & 1040 \\ \hline
2 & Project management & 448 & 7 & 38 & 0 & 493 \\ \hline
3 & Access management & 399 & 2 & 52 & 39 & 492 \\ \hline
4 & Vulnerability management & 336 & 3 & 99 & 2 & 440 \\ \hline
5 & Problem solving & 204 & 13 & 28 & 0 & 245 \\ \hline
\end{tabular}%
}
\end{table}

\smallbreak

Table \ref{hard_skills} presents the identified hard skills for cybersecurity professionals. Information security and information technology are identified as the most important hard skills. Information security mostly encapsulates skills required for ensuring the confidentiality, integrity, and availability of the data. Information technology is a generic skill that expects security professionals to have a good understanding of the technological landscape of their respective organizations. Interestingly, our results did not reveal cryptography as one of the key hard skills, unlike those mentioned in cyber forums such as CyberGeek \footnote{https://www.cyberseek.org/certifications.html}.
\begin{table*}[ht]
\caption{Identified hard skills for cybersecurity professionals. 'Nums' denote the number of job ads/posts mentioning the particular skill.}\label{hard_skills}
\scriptsize
\centering
\begin{tabular}{|l|l|l|l|l|l|l|}
\hline
\textbf{Rank} & \textbf{Skill} & \textbf{Dice} & \textbf{SA Job Search} & \textbf{Seek} & \textbf{Stack Overflow} & \textbf{Total} \\ \hline \hline
1 & Information security & 683 & 13 & 270 & 29 & 995 \\ \hline
2 & Information technology & 516 & 5 & 89 & 5 & 615 \\ \hline
3 & Security clearance & 481 & 8 & 72 & 2 & 563 \\ \hline
4 & Network security & 284 & 8 & 73 & 148 & 513 \\ \hline
5 & Incident response & 362 & 9 & 100 & 1 & 472 \\ \hline
6 & Information systems & 418 & 3 & 37 & 1 & 459 \\ \hline
7 & Security operations & 267 & 8 & 146 & 2 & 423 \\ \hline
8 & Security controls & 268 & 2 & 113 & 20 & 403 \\ \hline
9 & Application security & 180 & 3 & 44 & 138 & 365 \\ \hline
10 & Security services & 213 & 7 & 102 & 36 & 358 \\ \hline
11 & Cloud security & 241 & 3 & 57 & 46 & 347 \\ \hline
12 & Software development & 272 & 10 & 18 & 9 & 309 \\ \hline
13 & Security architecture & 212 & 3 & 48 & 29 & 292 \\ \hline
14 & Service delivery & 89 & 1 & 36 & 1 & 127 \\ \hline
15 & Enterprise architecture & 56 & 1 & 10 & 1 & 68 \\ \hline
16 & Internal audit & 37 & 0 & 6 & 1 & 44 \\ \hline
17 & Data architecture & 32 & 0 & 0 & 0 & 32 \\ \hline
18 & Digital technology & 10 & 0 & 0 & 4 & 14 \\ \hline
19 & Risk compliance & 3 & 0 & 0 & 0 & 3 \\ \hline
20 & Audit risk & 1 & 0 & 0 & 0 & 1 \\ \hline
\end{tabular}%
\end{table*}
\smallbreak

Table \ref{certifications} shows the most in-demand certifications for security professionals. We found that CISSP\footnote{https://www.isc2.org/Certifications/CISSP} and CISM\footnote{https://www.isaca.org/credentialing/cism} are the most mentioned certifications in the job ads. CISSP costs around \$749 and CISM costs around \$760. We also found that 50.2\% job ads expect some kind of professional certifications from the job seeker. 
\begin{table*} [ht]
\caption{Identified certifications for cybersecurity professionals. 'Nums' denote the number of job ads mentioning the particular certification}\label{certifications}
\centering
\scriptsize
\begin{tabular}{|l|l|l|l|l|l|l|l||l|}
\hline
\textbf{Rank} & \textbf{Dice} & \textbf{Nums} & \textbf{SA Job Search} & \textbf{Num} & \textbf{Seek} & \textbf{Num} & \textbf{Certification} & \textbf{Total} \\ \hline \hline
1 & CISSP & 907 & CISSP & 13 & CISSP & 270 & CISSP & 995 \\ \hline
2 & CISA & 304 & ITIL & 10 & CISM & 146 & CISM & 615 \\ \hline
3 & CISM & 292 & CISM & 9 & ITIL & 113 & CISA & 563 \\ \hline
4 & CEH & 279 & GIAC & 8 & CISA & 102 & CEH & 513 \\ \hline
5 & CompTIA & 219 & OSCP & 8 & CEH & 100 & CompTIA& 472 \\ \hline
6 & ITIL & 175 & CompTIA & 8 & OSCP & 89 & ITIL & 459 \\ \hline
7 & GIAC & 165 & CEH & 7 & GIAC & 73 & GIAC & 423 \\ \hline
8 & GCIH & 124 & CISA & 5 & CompTIA & 72 & GCIH & 403 \\ \hline
9 & CASP+ & 103 & CIPP & 3 & GSEC & 57 & OSCP & 365 \\ \hline
10 & OSCP & 94 & GSEC & 3 & SSCP & 48 & GSEC & 358 \\ \hline
11 & GSEC & 86 & SSCP & 3 & GCIH & 44 & CASP+ & 347 \\ \hline
12 & SSCP & 64 & CASP+ & 3 & CASP+ & 37 & SSCP & 309 \\ \hline
13 & CIPP & 13 & GCIH & 2 & CIPP & 36 & CIPP & 292 \\ \hline
\end{tabular}%
\end{table*}

\smallbreak
In order to provide a clear understanding of what the industry professionals understand by these hard skills, a detailed dictionary has been provided in \autoref{tab:skills_dictionary}. This information has been sourced from NIST (National Institute of Standards and Technology), SANS (SysAdmin, Audit, Network, Security), and ISO/IEC standards. Providing a dictionary for what these skills entail will be useful for  individuals looking to enter the field of cybersecurity or for educational institutions seeking to teach the necessary skills to their students.

\begin{tcolorbox}[left=0pt, top=0pt, right=0pt, bottom=0pt]
\textbf{Key insights for RQ1:} Combining the data in Stack Overflow and job advertisements, communication skills and project management are the skills that are in high demand in soft skills; information security, information technology and security clearance are in high demand in hard/technical skills. Regarding professional certifications, CISSP, CISA and CISM are in high demand.
\end{tcolorbox}
\begin{table*}[ht]
\centering
\tiny
\renewcommand{\arraystretch}{1.8} 
\setlength{\tabcolsep}{2pt} 
\caption{Hard Skills in Cybersecurity: A Dictionary}
\begin{tabularx}{\textwidth}{|p{2.5cm}|>{\raggedright\arraybackslash}X|X|X|X|}
\hline
\textbf{Skill} & \textbf{Definition} & \textbf{Key Components} & \textbf{Common Tools} & \textbf{Primary Source} \\
\hline
Information Security & Protection of information using a risk management approach to maintain Confidentiality, Integrity, and Availability. & Risk Assessment, Security Controls Implementation, Security Policy Development, Security Awareness Training. & SIEM Tools, DLP Solutions, Encryption Tools & NIST SP 800-12 \\
\hline
Information Technology & Governance of IT systems for efficient use of information and technology  & Infrastructure Management, Systems Administration, Network Management, Technical Support. & System monitoring tools, Service desk software, Network management systems. & ISO/IEC 38500 \\
\hline
Security Clearance & Authorization to classified information or restricted areas based on background checks. & Background Investigation, Continuous Monitoring, Access Level Management. & Personnel Security Systems, Security Management Platforms. & NIST SP 800-181 \\
\hline
Network Security & Protection of network infrastructure and data transmitted through networks. & Perimeter Defense, Network Monitoring, Access Control, Intrusion Detection. & Firewalls, IDS/IPS Systems, VPNs & ISO/IEC 27033 \\
\hline
Incident Response & Organized approach to addressing and managing security breaches or cyberattacks. & Incident Response Lifecycle & EDR platforms, SOAR tools, Digital Forensics Tools. & NIST SP 800-61 \\
\hline
Information Systems & Integrated set of components for collecting, storing, processing, and providing information. & Systems Analysis, Database Management, Software Integration. & ERP Systems, Database Management Systems. & ISO/IEC 25010 \\
\hline
Security Operations & Day-to-day activities to monitor, detect, and respond to security events. & Security Monitoring, Alert Management, Threat Hunting. & SIEM Tools, Threat Intelligence Platforms  & SANS SOC Guidelines \\
\hline
Security Controls & Safeguards or countermeasures to avoid, detect, counteract, or minimize security risks. & Technical Controls, Administrative Controls. & Access Control Systems, Firewalls. & NIST SP 800-53 \\
\hline
Application Security & Measures taken to improve security of applications by finding, fixing, and preventing vulnerabilities. & Secure SDLC, Vulnerability Assessment, Security Testing. & SAST/DAST Tools, WAF, Code Analysis Tools. & SANS AppSec \\
\hline
Security Services & Organizational and technical services that implement and operate security controls. & Security Consulting, Managed Security, Security Assessment. & GRC Platforms, Assessment Tools. & ISO/IEC 27001 \\
\hline
Cloud Security & Protection of data, applications, and infrastructure associated with cloud computing. & Cloud Configuration, Identity Management, Data Protection. & CASB, Cloud Security Tools. & NIST SP 800-144 \\
\hline
Software Development & Designing, programming and maintaining software & Software Implementation. & Version Control Systems, CI/CD Tools. & ISO/IEC 12207 \\
\hline
Security Architecture & Design of an organization's security infrastructure and policies. & Security Framework Design, Risk Assessment, Control Selection. & Architecture Tools, Security Design Tools. & NIST SP 800-160 \\
\hline
Service Delivery & Management and delivery of IT services to meet business requirements. & Service Level Management, Capacity Management, Availability Management. & Service Desk Software, ITSM Platforms. & ISO/IEC 20000 \\
\hline
Enterprise Architecture & Strategic approach to linking business structure and processes with IT infrastructure. & Business Architecture, Information Architecture, Technology Architecture. & EA Modeling Tools, Process Mapping Tools. & NIST EA Framework \\
\hline
Internal Audit & Independent evaluation of organization's operations, controls, and compliance. & Risk Assessment, Control Testing, Compliance Review. & Audit Management Software, Data Analytics Tools. & ISO 19011 \\
\hline
Data Architecture & Designing systems for collecting, storing, and delivering data. & Data Modeling, Database Design, Data Governance. & Data Modeling Tools, Database Systems. & NIST SP 800-122 \\
\hline
Digital Technology & Using electronic tools to process and store data. & Digital Infrastructure, Digital Platforms, Digital Solutions. & Digital Platform Tools, Digital Analytics Tools. & ISO/IEC 2382 \\
\hline
Risk Compliance & Adherence to laws, standards, policies and procedures & Compliance Monitoring, Risk Assessment, Control Implementation. & GRC Platforms, Risk Assessment Tools. & NIST RMF \\
\hline
Audit Risk & Auditing for errors or irregularities & Inherent Risk, Control Risk, Detection Risk. & Audit Software, Risk Analysis Tools. & ISO 19011 \\
\hline
\end{tabularx}
\label{tab:skills_dictionary}
\end{table*}

\subsection{RQ2: Skills and type of security positions}
Table \ref{soft_skills_roles}, Table \ref{hard_skills_roles}, and Table \ref{certifications_roles} respectively show the top soft skills, hard skills, and certifications with respect to various cybersecurity roles. In these tables, \% specifies the percentage of job ads for a specific role that requires the respective skill. For example, 140/182 (77\%) job ads for a tester role mention communication as an important skill. We can observe that the required skills vary across roles. For instance, project management is considered an important skill for consultants but not for testers. Communication is considered an important skill for almost all cyber roles. 
\begin{table*}[ht]
	\caption{Identified soft skills for various cybersecurity roles (job ads data)}\label{soft_skills_roles}
	\centering
	\tiny
	\begin{tabular}{|p{2.8cm} |p{1cm} |p{1.1cm} |p{1cm} |p{1.2cm} |p{1cm} |p{1cm} |p{1cm} |p{1.2cm}|}
		\hline
		\textbf{Soft Skills}    &\textbf{Analyst} &\textbf{Architect} &\textbf{Auditor}   &\textbf{Consultant}    &\textbf{Engineer}  &\textbf{Manager}   &\textbf{Tester}    &\textbf{Specialist}\\\hline\hline
	    {Communication skills}        &321(68\%)    &123(64\%)    &394(42\%)    &419(61\%)    &382(16\%)    &322(49\%)    &140(77\%)    &230(42\%)\\
		Problem-solving skills      &71(15\%)    &0    &0    &0    &0    &0    &0    &0\\ 
		Project management          &0    &0    &107(2\%)    &93(13\%)    &149(28\%)    &128(19\%)    &0    &89(6\%)\\
		Management skills           &0    &0    &89(10\%)    &0    &0    &0    &0    &0\\ 
		Vulnerability management    &0    &0    &0    &0    &0    &0    &42(23\%)    &63(12\%)\\\hline
\end{tabular}
\end{table*}
\smallbreak
The same is true for hard skills, where a particular skill might be considered important for one role but not for another. As an example, enterprise architecture is an important skill for architects but not so much important for an analyst. Information Security emerges as the most essential skill across multiple roles including for security analysts, consultants , engineers , managers , testers  and specialists. This is because foundational knowledge of information security is essential to protect systems, advise people on secure practices, develop robust infrastructures, manage organizational security strategies, and identify vulnerabilities. Security Architecture is the the most essential skill for architects since their job revolves around designing secure systems that are both scalable and resilient. Digital Technology is most important for auditors. This is in line with their focus on evaluating digital systems, ensuring compliance with standards, and identifying gaps in an organization’s digital framework. Knowledge of security operations plays an important role for security engineers, testers, and specialists, highlighting the importance of implementing and maintaining effective security measures within systems.  Knowledge of internal audit procedures is mostly required for auditors helping them assess internal processes and controls to ensure everything stays secure.
\begin{table*}[ht]
\centering
	\tiny
	\caption{Identified hard skills for various cybersecurity roles (job ads data)}\label{hard_skills_roles}
	\begin{tabular}{|p{2.5cm} |p{1cm} |p{1.2cm} |p{1cm} |p{1.3cm} |p{1cm} |p{1cm} |p{1cm} |p{1.2cm}|}
		\hline  
		\textbf{Technical Skills}    &\textbf{Analyst} &\textbf{Architect} &\textbf{Auditor}   &\textbf{Consultant}    &\textbf{Engineer}  &\textbf{Manager}   &\textbf{Tester}    &\textbf{Specialist}\\\hline\hline
		Information security    &{458(42\%)}  &{204(26\%)}   &217(18\%)   &{551(39\%)}    &{480(31\%)}    &{549(44\%)}    &{553(31\%)}    &{346(42\%)}\\
		Information technology  &123(11\%)  &0          &255(22\%)  &0    &185(12\%)    &165(13\%)    &0    &106(13\%)\\
		Incident response       &115(10\%)  &0          &0          &0    &143(9\%)    &88(7\%)    &290(17\%)    &89(11\%)\\
		Information systems     &0          &0          &109(9\%) &0    &0    &0    &0    &0\\ 
		Cloud security          &0          &48(6\%)  &0          &0    &0    &0    &0    &0\\
		Security operations     &80(7\%)    &0          &0          &140(10\%)    &162(10\%)    &131(10\%)    &187(11\%)    &145(17\%)\\
		Application security    &0          &0          &0          &0    &0    &0    &163(19\%)    &0\\
		Security controls       &122(11\%)  &40(5\%)    &0          &96(17\%)    &146(9\%)    &0    &143(9\%)    &64(8\%)\\
		Security architecture   &119(11\%)  &171(22\%)  &0          &0    &0    &0    &0    &0\\
		Network security        &89(8\%)    &46(6\%)   &0          &0    &172(11\%)    &0    &0    &0\\
		Enterprise architecture &0          &174(23\%) &0          &0    &0    &0    &0    &0\\ 
		Software development    &0          &51(7\%)  &0          &0    &0    &0    &0    &0\\ 
		Digital technology      &0          &40(5\%)  &0          &0    &0    &0    &0    &0\\
		Internal audit          &0          &0          &{423(36\%)} &0    &0    &0    &0    &0\\
		Risk compliance         &0          &0          &89(8\%)   &0    &0    &91(7\%)    &0    &0\\
		Audit risk              &0          &0          &83(7\%)  &0    &0    &0    &0    &0\\
		Data architecture       &0          &0          &0          &196(14\%)    &0    &0    &0    &0\\ 
		Teaching programs       &0          &0          &0          &196(14\%)    &0    &0    &0    &0\\
		Security service        &0          &0          &0          &118(8\%)    &0    &107(9\%)    &0    &79(8\%)\\
		Diversity inclusion     &0          &0          &0          &98(7\%)    &0    &0    &0    &0\\
		Security clearance      &0          &0          &0          &0    &109(7\%)    &0    &0    &0\\
		Service delivery        &0          &0          &0          &0    &0    &126(10\%)    &0    &0\\ 
		Penetration testing     &0          &0          &0          &0    &0    &0    &220(13\%)    &0\\
    	Technical security      &0          &0          &0          &0    &0    &0    &120(7\%)    &0\\
		Security testing        &0          &0          &0          &0    &0    &0    &80(5\%)    &0\\
		Security systems        &0          &0          &0          &0    &175(11\%)    &0    &0    &0\\\hline
\end{tabular}

\end{table*}
\smallbreak 
With respect to certifications, we found that certifications are also common among various roles. This means that the majority of these certifications, especially CISSP, CISA, and CISM, are generic and helpful for acquiring various several roles. 
 \begin{table*}[ht]
	\caption{Identified certifications for various cybersecurity roles (job ads data)}\label{certifications_roles}
	\centering
	\tiny
	\begin{tabular}{|p{1.5cm} |p{1cm} |p{1.3cm} |p{1cm} |p{1.3cm} |p{1.3cm} |p{1.2cm} |p{1cm} |p{1.3cm}|}
		\hline 
  
  \textbf{Certification}    &\textbf{Analyst} &\textbf{Architect} &\textbf{Auditor}   &\textbf{Consultant}    &\textbf{Engineer}  &\textbf{Manager}   &\textbf{Tester}    &\textbf{Specialist}\\\hline\hline
		CISSP   &{87(26\%)}  &{39(31\%)}  &32(17\%)                   &{109(30\%)} &{147(30\%)} &{91(24\%)}  &{136(25\%)}    &{68(26\%)}\\
		CISA    &21(6\%)                    &8(6\%)                     &{55(30\%)}  &45(12\%)                   &43(9\%)                    &52(14\%)       &48(9\%)    &39(15\%)\\
		CISM    &34(10\%)                   &18(15\%)                   &38(21\%)                   &70(19\%)                   &54(11\%)                   &78(21\%)       &86(15\%) &52(20\%)\\
		GIAC    &24(7\%)                    &11(9\%)                    &0                          &23(6\%)                    &31(6\%)                    &20(5\%)        &33(6\%) &13(5\%)\\
		CompTIA &42(13\%)                   &6(5\%)                     &2(1\%)                     &6(2\%)                     &21(4\%)                    &13(3\%)        &28(5\%) &9(3\%)\\
		ITIL    &22(7\%)                    &16(13\%)                   &46(25\%)                   &61(17\%)                   &63(13\%)                   &74(20\%)       &55(10\%) &45(17\%)\\
		CEH     &38(12\%)                   &8(6\%)                     &9(5\%)                     &19(5\%)                    &57(11\%)                   &21(6\%)        &42(8\%) &20(7\%)\\
		GCIH    &12(4\%)                    &3(2\%)                     &0                          &4(1\%)                     &13(2\%)                    &2(1\%)         &8(1\%)  &4(1\%) \\
		CASP+   &2(1\%)                     &1(1\%)                     &0                          &0                          &5(1\%)                     &0              &2(0\%)  &2(1\%) \\
		OSCP    &13(4\%)                    &7(6\%)                     &1(1\%)                     &11(3\%)                    &29(6\%)                    &8(2\%)         &81(15\%) &9(3\%)\\
		GSEC    &17(5\%)                    &6(5\%)                     &0                          &10(3\%)                    &18(4\%)                    &7(2\%)         &24(4\%) &2(1\%)\\
		SSCP    &16(5\%)                    &1(1\%)                     &1(1\%)                     &4(1\%)                     &13(3\%)                    &5(1\%)         &11(2\%) &2(1\%)\\
		CIPP    &1(0\%)                     &0                          &0                          &2(1\%)                     &2(0\%)                     &2(1\%)         &1(0\%)  &1(0\%)\\\hline
	\end{tabular}
\end{table*}

\begin{tcolorbox}[left=0pt, top=0pt, right=0pt, bottom=0pt]
\textbf{Key insights for RQ2:} The required skills vary across various cybersecurity roles. Except for communication skills, no other skill is pervasively applicable to all roles. 
\end{tcolorbox}

\subsection{RQ3: Programming languages required}

\begin{figure}[ht]
 \centering
 \makebox[\textwidth][c]{\includegraphics[width=8 cm, height=4.5cm]{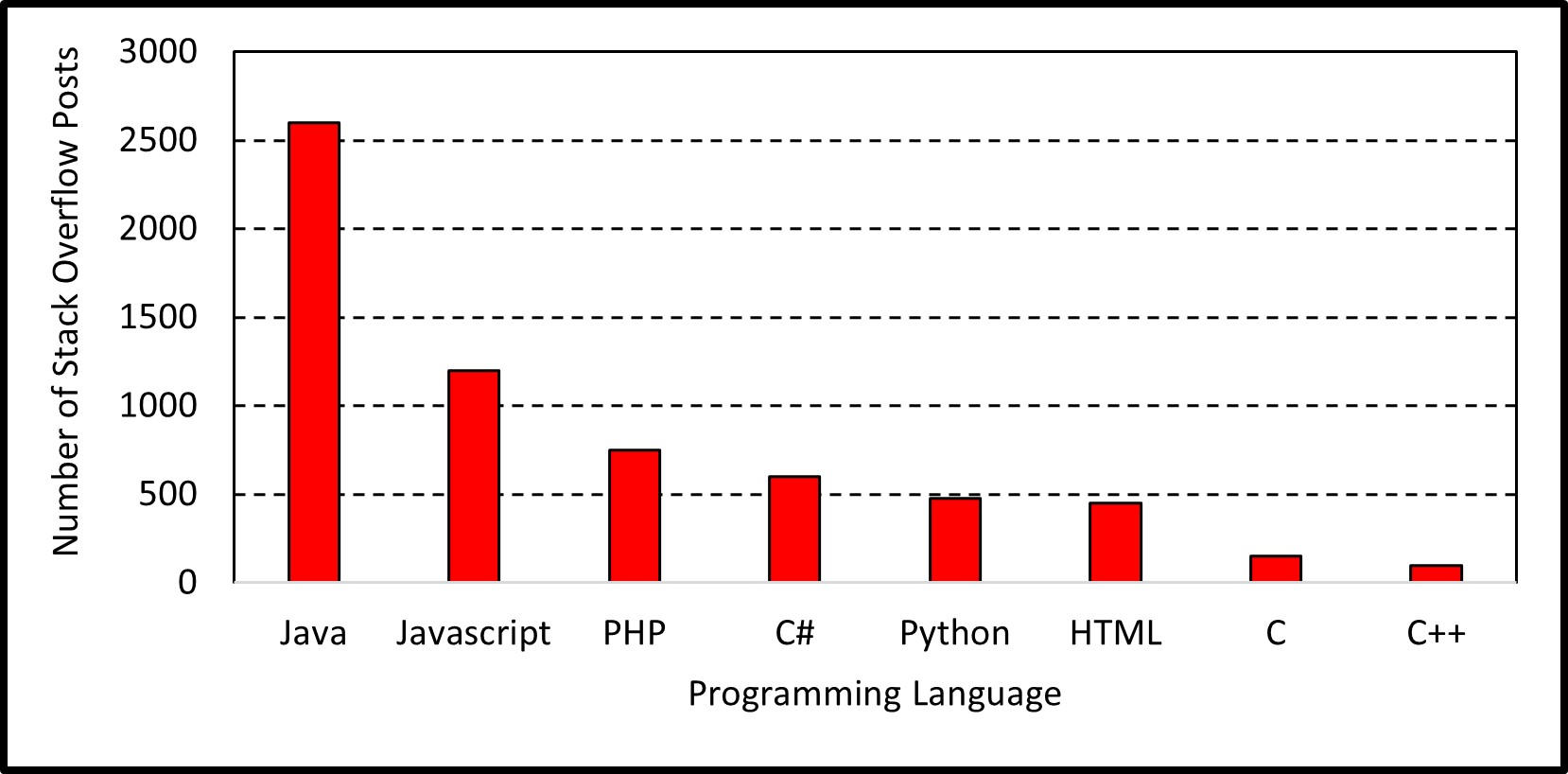}}%
 \caption{Number of cybersecurity posts for programming languages on Stack Overflow}
 \label{rq1-3}
 \vspace{-1\baselineskip}
\end{figure}

\begin{figure}[t]
 \centering
 \makebox[\textwidth][c]{\includegraphics[width=12.5cm, height=8cm]{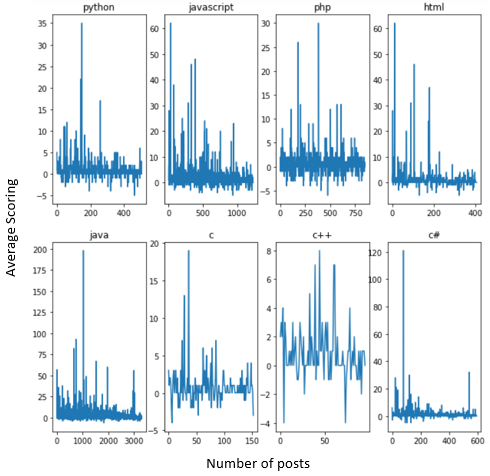}}%
 \caption{Programming languages scores of cybersecurity-related posts on Stack Overflow.}
 \label{rq1-5}
 \vspace{-1\baselineskip}
\end{figure}

We answer this question based on the data collected from Stack Overflow data. As shown in Figure \ref{rq1-3}, we found that programming languages, including C, C++, C\#, HTML, Java, JavaScript, PHP, and Python, are essential skills for security professionals. Java is the most mentioned programming language. Since each question from Stack Overflow contains multiple tags, a tag may appear in different questions. Therefore, we calculate the average score of each tag. As shown in Figure \ref{rq1-5}, the average score of Java is the highest. Java is popular in the cybersecurity community due to its cross-platform compatibility, large community support, and robust libraries. In contrast, it intuitively shows which programming languages have higher scores and are more suitable for use in the cybersecurity industry. For example, the highest number of questions in security-related posts on Stack Overflow were about Java. Hence, Java has a higher score as shown in Figure \ref{rq1-5}. C++, on the other hand, has less number of questions and consequently scored significantly lower.

\begin{tcolorbox}[left=0pt, top=0pt, right=0pt, bottom=0pt]
\textbf{Key insights for RQ3:} Security professionals need several programming languages. However, Java is the most commonly used programming language in cyber projects.
\end{tcolorbox}

\section{Discussion}\label{D}
In this section, we compare the findings from job ads and Stack Overflow. We also discuss the implications of our findings and the threats to the validity of our findings. 

\subsection{Comparing cyber skills in Job Ads and Stack Overflow}
We compared the skills identified from the two sources - job ads and Stack Overflow. We found that Stack Overflow posts related to cybersecurity contain more technical terms such as frameworks, tools, and programming languages. These details are often missing in job ads. On the other hand, we observed that Stack Overflow data contains very little information related to soft skills such as communication and project management. In contrast, job ads contain plenty of requirements related to soft skills. We also compared the skills mentioned in job ads to the ones highlighted as important by CyberSeek \footnote{https://www.cyberseek.org/index.html}. We found that the majority of the technical skills mentioned in job ads are the same as those deemed important by CyberSeek. Furthermore, we found that CyberSeek emphasizes the importance of certifications for certain roles such as architects and consultants. However, job ads mention these certifications irrespective of the advertised role in the majority of the cases. 

\subsection{Implications of our findings}

\textit{Cybersecurity personnel.} Our findings are useful for individuals aiming to enter the cybersecurity industry. Such individuals can learn what roles exist in cybersecurity and what are soft and technical skills expected for effectively performing in such roles. Further, individuals can learn about the most popular certifications. \textit{Cybersecurity ad design} Our findings can serve as a guide for companies to design better job ads/descriptions. Companies can include the skills identified in our study. Moreover, our findings can help companies to accurately understand the needs of internal personnel, reduce additional training, and improve the efficiency of employees. \textit{Education and Research.} In addition to industry, our findings are also useful for educational institutes. Given that there is an exponential surge in cybersecurity degree offerings, institutes can learn about the in-demand cyber skills in the industry. Accordingly, they can design their curriculum to teach the most in-demand cyber skills. 

\subsection{Threats to Validity}
Identifying and extracting relevant questions and answers from Stack Overflow through cybersecurity labels alone threatens the completeness of our data. To reduce this threat, we used sets of job titles, skills, and certification tags based on job breakdown and skill requirements provided in CyberSeek. Not all jobs are advertised through recruitment portals. For example, appointments for senior positions are often based on a professional network. Similarly, fresh graduates are hired via internships and campus recruitment too. To mitigate this threat, unlike previous studies \cite{2,3,5,10,12,15,29,caulkins2019cybersecurity,jerman2022cybersecurity,peslak2019cybersecurity}, we collected 12,161 job ads from three different sources. In addition, the timeliness of job advertisements also poses a specific threat. This is because the irregular posting and closing of job advertisements have prevented us from studying the evolution of recruitment demand in recent years.Although we collected a large number of job ads and Stack Overflow posts, we still cannot claim the generalizability of our findings. To mitigate this threat, we first studied 1000 job ads. We observed that adding more ads changes the frequency distributions but does not change the skill set significantly. 
\smallbreak
While this study provides insights based on job advertisements and Stack Overflow data, future research could focus on the direct involvement of active professionals in the field. A survey targeting cybersecurity practitioners should be conducted to validate and complement the findings of this study. Such a survey could include questions aimed at understanding the most essential technical skills, the certifications considered most beneficial, the programming languages frequently used in practice, and the soft skills critical for daily responsibilities. Additionally, the survey could also cover emerging trends or overlooked competencies that may not be prominent in job advertisements or online forums. This would help bridge the gap between theoretical and practical workplace requirements and enhance the overall relevance and utility of our findings.

\section{Conclusion}\label{C}
In this study, we analyzed data from 12,161 cybersecurity job ads and 49,002 Stack Overflow posts on cybersecurity to determine the most in-demand skills, the relation between cyber skills and cyber roles, and the programming languages critical for security professionals. The results of the study show that communication skills and project management skills are in high demand for soft skills, while information security, information technology, and security clearance are in high demand for technical skills. Except for communication skills, the demand for cyber skills varies across various cyber roles. The study also found that Java is the most commonly used programming language.

In terms of future work, the study provides a foundation for further research on the changing demands for cybersecurity skills. For example, it would be interesting to see how the demand for specific skills changes over time and how the skills in high demand vary across different industries and regions. Additionally, mapping certifications to career tracks could bridge the gap between qualifications and industry needs. Future studies could also explore the relationship between specific skills and job performance, as well as the return on investment for acquiring specific skills.


\bibliography{CRP}
\end{document}